\documentclass[namedreferences]{solarphysics}

\usepackage[hyperref,optionalrh,solaromanenum]{spr-sola-addons} 

\usepackage{epsfig}                     
\usepackage{epstopdf} 
\usepackage{graphicx}                    
\usepackage{color}                       
\usepackage{breakurl}                         

\chardef\us=`\_

\begin{document}

\begin{article}

\begin{opening}

\title{Relationship of type III radio bursts with quasi-periodic pulsations in a solar flare}

\author[addressref={aff1,aff2},corref,email={elena.kupriyanova@kuleuven.be}]{\inits{}\fnm{E. G.}~\lnm{Kupriyanova}}

\author[addressref=aff3,email={lkk@iszf.irk.ru}]{\inits{}\fnm{L. K.}~\lnm{Kashapova}} 

\author[addressref=aff4,email={hamish.reid@glasgow.ac.uk}]{\inits{}\fnm{H. A. S.}~\lnm{Reid}} 

\author[addressref=aff5,email={irina@srd.sinp.msu.ru}]{\inits{}\fnm{I. N.}~\lnm{Myagkova}} 

\address[id=aff1]{KU~Leuven, Departement Wiskunde, Celestijnenlaan 200B bus 2400, BE-3001 Leuven, Belgium}
\address[id=aff2]{Central Astronomical Observatory at Pulkovo of the RAS, 
Pulkovskoe shosse 65, Saint-Petersburg, 196140 Russia}
\address[id=aff3]{Institute of Solar-Terrestrial Physics SB RAS, Lermontova st. 126A, 664033 Irkutsk, Russia}
\address[id=aff4]{SUPA School of Physics and Astronomy, University of Glasgow, Glasgow G12 8QQ, UK}
\address[id=aff5]{Skobeltsyn Institute of Nuclear Physics, Lomonosov Moscow State University, Leninskie gory st. 1c2, 119991 Moscow, Russia}


\runningauthor{Kupriyanova et al.}
\runningtitle{Relationship of type III radio bursts with quasi-periodic pulsations in a solar flare}

\begin{abstract}
We studied a solar flare with pronounced quasi-periodic pulsations detected in the microwave, X-ray, and radio bands. We used the methods of correlation, Fourier, and wavelet analyses to examine the temporal fine structures and relationships between the time profiles in each wave band. We found that the time profiles of the microwaves, hard X-rays and type III radio bursts vary quasi-periodically with the common period of 40--50~s.  The average amplitude of the variations is high, above 30\% of the background flux level and reaching 80\% after the flare maximum.  We did not find the periodicity in either the thermal X-ray flux component or source size dynamics.  Our findings indicate that the detected periodicity is likely to be associated with periodic dynamics in the injection of non-thermal electrons, that can be produced by periodic modulation of magnetic reconnection.
\end{abstract}


\keywords{Corona, Radio Emission; Flares, Impulsive Phase; Flares, Energetic Particles; Magnetic Reconnection, Observational Signatures; Radio Bursts,  Type III; Radio Bursts, Microwave (mm, cm); X-Ray Bursts, Association with Flares}

\end{opening}


\section{Introduction}\label{s:Introduction}

Quasi-periodic pulsations (QPPs) are frequently seen in the time profiles of  solar and stellar flare emission
\citep[see reviews][]{2009SSRv..149..119N, 2016SSRv..200...75N}. Among all the possibilities  of QPPs explanation, two reasons are discussed more actively: magnetohydrodynamic (MHD) oscillations in plasma waveguides and oscillations in current sheets. 

MHD oscillations in plasma waveguides observed in the microwave and X-ray bands cover the wide range of time scales, from a few  seconds \citep{1969ApJ...155L.117P, 1983ApJ...271..376K, 2001ApJ...562L.103A, 2003ApJ...588.1163G, 2005A&A...439..727M, 2008ApJ...684.1433F, 2009A&A...493..259I, 2013SoPh..284..559K} 
up to tens of minutes \citep{2006A&A...460..865M,2014SoPh..289.1239N, 2014ARep...58..573K}. Microwave and X-ray emissions are 
usually related to energetic electrons accelerated during the flare. Gyrating around the magnetic field lines the relativistic electrons may emit via the gyrosynchrotron mechanism. At the same time electrons precipitate into loop footpoints where they produce hard X-ray (HXR) emission via bremsstrahlung \citep{1982SvAL....8..132Z}. 
MHD wave processes in coronal magnetic structures can have an effect on the process of electron acceleration, or on the emission of already accelerated electrons, leading to quasi-periodicities in, \textit{e.g.}, microwave and in X-ray emissions.

Oscillations in current sheets are usually connected with series of reconnection \citep[see][ for review]{2009SSRv..149..119N}.
When we are speaking about the reconnections and type~III bursts associated with them, we mean the characteristic times of a few seconds \citep{1994ApJ...431..432A, 2005A&A...437..691N, 2013A&A...555A..40A}.  \citet{2009A&A...494..329M}  
found oscillations with periods from one up to few minutes as a result of reconnection driven by simulations of magnetic flux emerging in a coronal hole.
\citet{2012ApJ...749...30M} 
studied the analogous model and found that the physical mechanism of oscillatory reconnection could generate quasi-periodic vertical outflows. Periods of 1.75--3.5~minutes could be obtained by varying the magnetic strength of the buoyant flux tube.
\citet{2015ApJ...804....4K} found observational evidence of quasi-periodic reconnection with the period near 3.3~minutes.
Repetitive reconnection was found to occur near footpoints of a loop arcade where the magnetic field had a fan configuration with a null-point \citep[][]{1998ApJ...502L.181A}. 
Consecutive reconnections
could cause the periodic acceleration of non-thermal electrons which propagated to the opposite footpoint along the arcade appearing there as repetitive increases in the brightness of X-ray and ultra violet emission.

Type~III radio bursts are created by accelerated electron beams resonantly exciting Langmuir waves \citep[see \textit{e.g.}][]{1973SoPh...31..207J, 2014RAA....14..773R}. The radio waves of type~III bursts are connected to a non-linear transformation of the electrostatic Langmuir waves to electromagnetic waves. The scattering of plasma waves on the ions of the background plasma leads to electromagnetic emission in the vicinity of the plasma frequency. The non-linear interaction of two plasma waves produces the emission around the second harmonic of the plasma frequency.  \citet{2015ApJ...807...72L} detected a similar periodicity in X-rays, emission of Extreme ultraviolet (EUV) lines, and in radio (several MHz) emission. They found that the peaks of the type~III burst corresponded to peaks of hard X-ray and extreme ultraviolet emissions.
The mechanism for these correlations is still not clear. However, close agreement between properties of the QPPs and type III bursts could be an indicator that QPPs are caused by current sheet oscillations.

\citet[][]{2016AdSpR..57.1456K} (hereafter Paper~1) revealed high-amplitude quasi-periodic pulsations during the impulsive phase of a solar flare on May 6, 2005.  The QPPs were co-phased in microwave and HXR emission.  However, an unambiguous conclusion about the origins of the observed QPPs was not reached.  The study was carried out based on the 
hard X-ray and microwave data
and did not consider type III radio bursts. The aim of this study is to reveal the most probable scenario of the observed QPPs based on analysis of the flare impulsive phase using additional data at microwave, X-ray, and radio wavelengths. As the spatial structure of the sources of microwaves and X-rays was analyzed in detail in Paper~1, we consider here the time behavior of the emissions and the derived plasma parameters.

The paper is organized as follows. Section~\ref{s:Observations} contains an overview of the flare observations used in the current study. Analyses of the time profiles are summarized in Section~\ref{s:Timeprofs}.  The periodic properties of the time series are analyzed using withe the use of the wavelet transform and the results are summarised in Section~\ref{s:QPP}. Comparison and discussion of arguments \textit{pro} and \textit{contra} of different interpretations are presented in Section~\ref{s:Discussion} and our conclusion and final remarks are in Section~\ref{s:Conclusion}.

\section{Observations}\label{s:Observations}
A C9.3 GOES class solar flare occurred on May 6, 2005 near the western limb with heliographic coordinates S05W67. In this study, we consider the impulsive phase that occurred between 03:07:00~UT to 03:11:00~UT. 

\subsection{X-ray data}\label{s:Xrays}
   \begin{figure}    
   \centerline{
\includegraphics[width=0.5\textwidth,clip=]{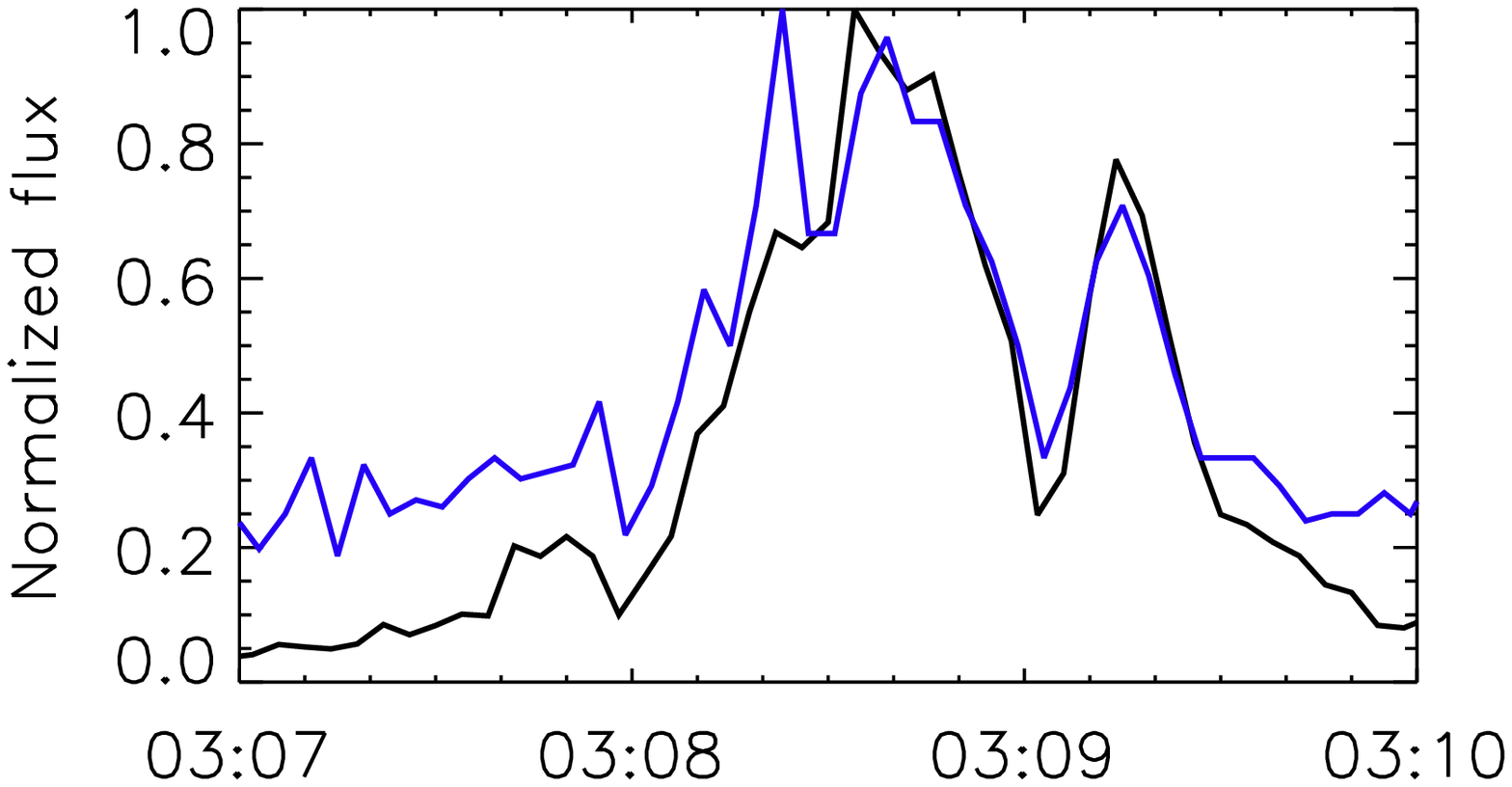}
\includegraphics[width=0.5\textwidth,clip=]{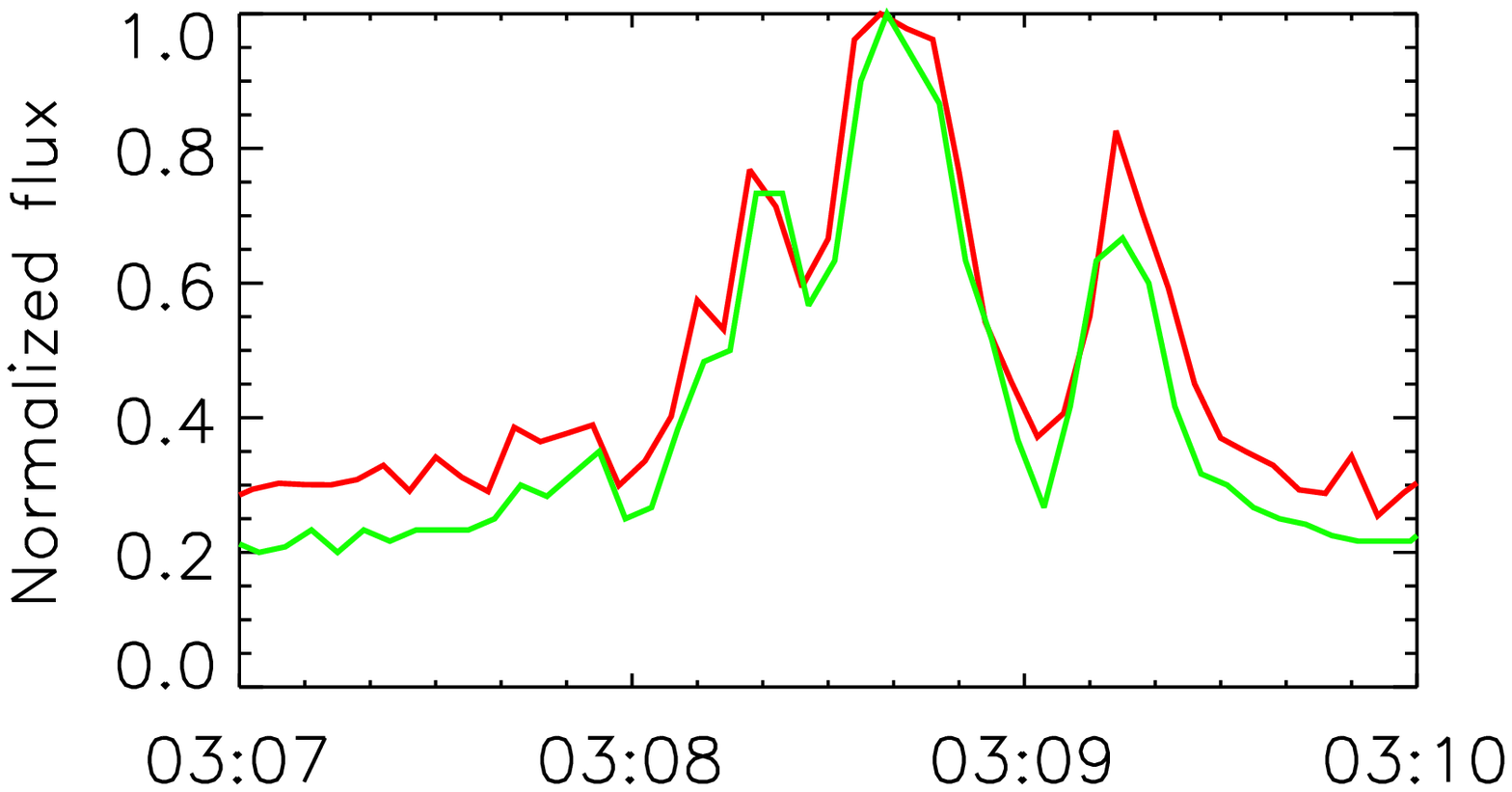}
              }
              \caption{
Time profiles of hard X-ray emission during the impulsive phase of the flare on May 6, 2005 normalised by the maxima. .
(a)  \textit{RHESSI} fluxes at 50--100~keV (black line) and \textit{SONG} fluxes at 43--82~keV (blue) 
(b) \textit{RHESSI} fluxes at 100--300~keV (red) and \textit{SONG} fluxes at 82--230~keV (green). }
   \label{f:SONG_RHESSI}
   \end{figure}
We used simultaneous hard X-ray observations of the flare by the 
{\it Reuven Ramaty High Energy Solar Spectroscopic Imager} (RHESSI) 
\citep[][]{2002SoPh..210....3L} and the {\it Solar Neutron and Gamma rays} (SONG) experiment onboard the three-axis stabilized {\it Complex Orbital Observations in the Near-Earth space of the Activity of the Sun} (CORONAS-F) solar space observatory \citep[][]{podorolsky2004first1466063}. 
These instruments together provide the 4-s time resolution measurements 
of X-rays and gamma rays in the range 3 keV to 200 MeV.  We use the 1D time profiles obtained by two different instruments to check for the possible effect and/or presence of artifacts in the QPPs observations \citep{2011A&A...530A..47I}.

The time profiles of \textit{RHESSI} fluxes at 50--100~keV and 100--300~keV, and the SONG fluxes at 43--82~keV and 82--230~keV are shown in Figure~\ref{f:SONG_RHESSI}. It is clear that the time profiles correspond to each other for the related energy bands and hence any artificial effects in the HXR time profiles could be excluded.

\subsection{Microwaves and radio waves}\label{s:MWRadio}
 \begin{figure}    
   \centerline{\includegraphics[width=0.75\textwidth,clip=]{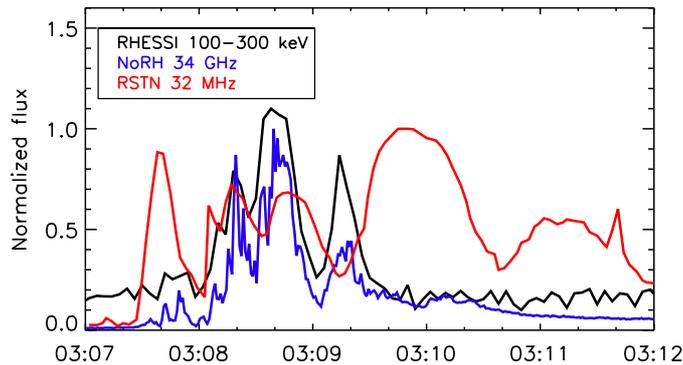}
              }
              \caption{
Time profiles (in UT) of NoRH, RHESSI, and RSTN Learmonth fluxes normalised by the maxima. RHESSI time profile is shifted artificially upward by 0.1 of arbitary unit in order to distinguish the time profiles of X-ray emission and NoRH flux.}
   \label{f:5}
   \end{figure}
 \begin{figure}
   \centerline{
\includegraphics[width=0.48\textwidth,clip=]{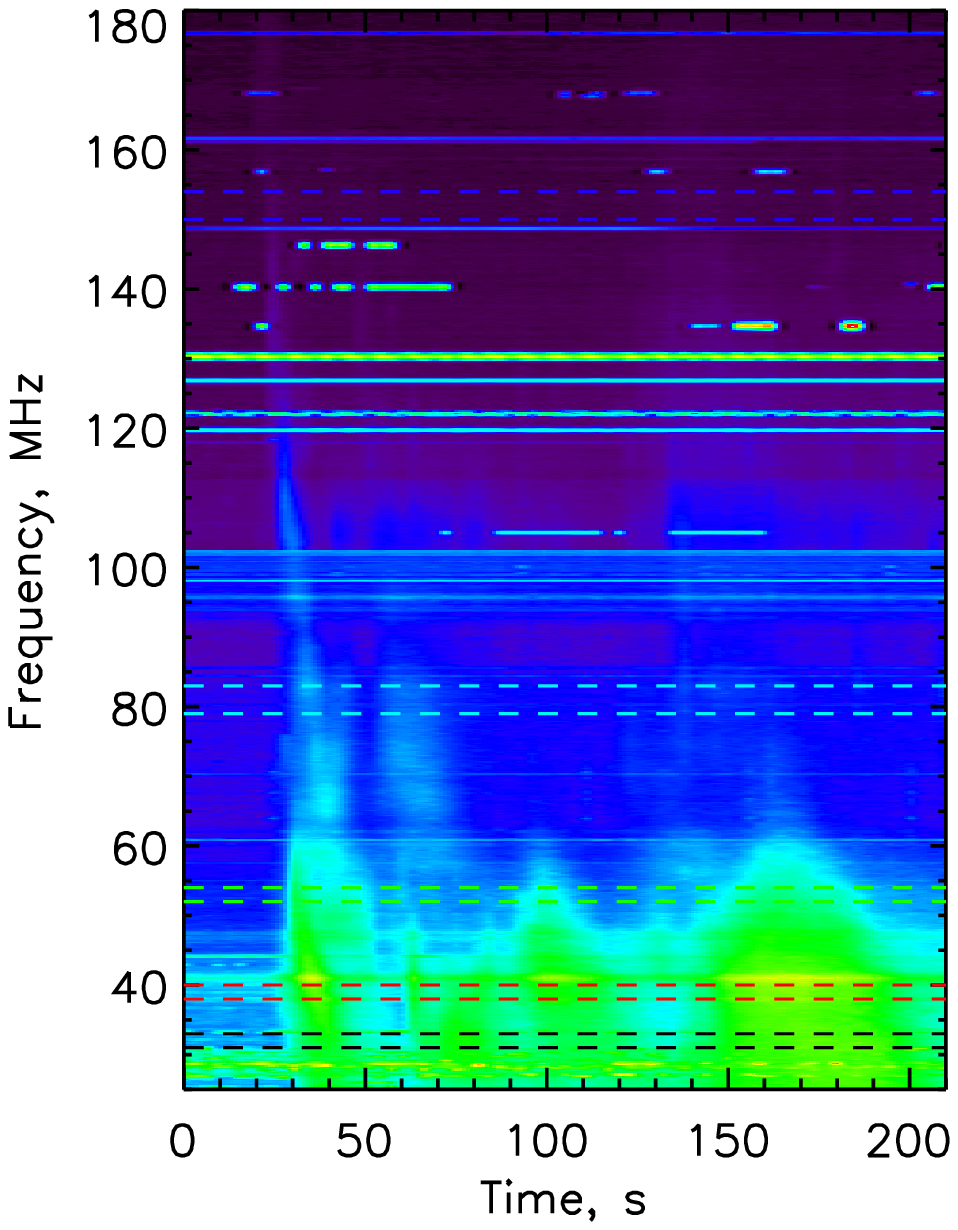}
\includegraphics[width=0.48\textwidth,clip=]{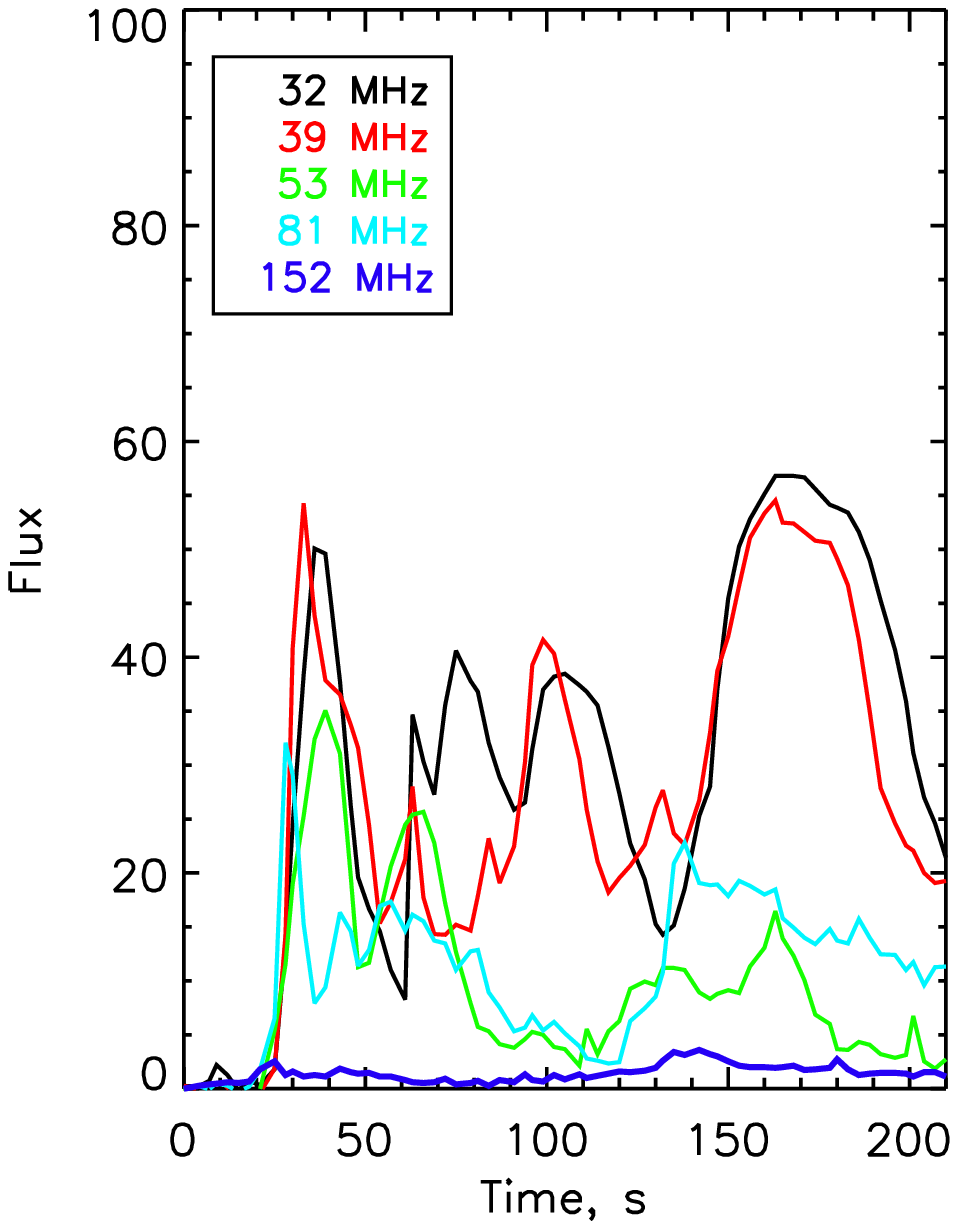}
}
   \caption{
		Left: dynamic spectrum for type~III bursts obtained with {\it the Learmonth Solar Radio Spectrograph}.
The horizontal axis is the time (in seconds) strating from 03:07:02~UT, 
the vertical axis is frequency. 
Five selected frequency bands are enclosed between the pairs of similarly colored dashed horizontal lines.
Right: the time profiles of the signals in these five spectral bands. 
The color of a curve corresponds to the color of the boundaries of the frequency bands shown in the left panel.
   }
  \label{f:RSTN}
 \end{figure}
We used the data obtained by the {\it Nobeyama Radioheliograph} (NoRH) and by the {\it Nobeyama Radio Polarimeters} (NoRP), Japan, \citep{1985PASJ...37..163N, 1994IEEEP..82..705N}  for the analysis of the microwave emission. Both instruments have high time resolution $\Delta t = 1$~s. The NoRH integrated time profiles at the frequencies 17~GHz and 34~GHz are calculated by integration of the values over the map of the intensity distribution at each time. The maps of the intensity distribution at the frequencies 17~GHz and 34~GHz are synthesized using the Fujiki and Hanaoka routines, correspondingly 
(\texttt{http://solar.nro.nao.ac.jp/norh/doc/man\_v33e.pdf}).

The spectral properties and time evolution of the flaring radio emission from 25~MHz to 180~MHz are studied using data from the ground-based {\it Learmonth Solar Radio Spectrograph}, Australia, included in {\it The Radio Solar Telescope Network} (RSTN). The cadence of the data is  $\Delta t = 3$~s. We have checked the agreement between the timelines of Nobeyama Observatory data and data from the {\it Learmonth Solar Radio Spectrograph}. For this purpose, we used microwave data within 1--17~GHz range obtained by both observatories. 
The comparison of the time profiles obtained by both instruments at similar frequencies revealed the advanced time shift of {\it the  RSTN} data by 5~s relative to the NoRP data.  We assumed the Nobeyama observatory time scale as the standard and so we added the time shift of 5~s to the RSTN data.

\section{Analysis of time profiles}\label{s:Timeprofs}
 \begin{figure}    
   \centerline{
      \includegraphics[width=0.48\textwidth,clip=]{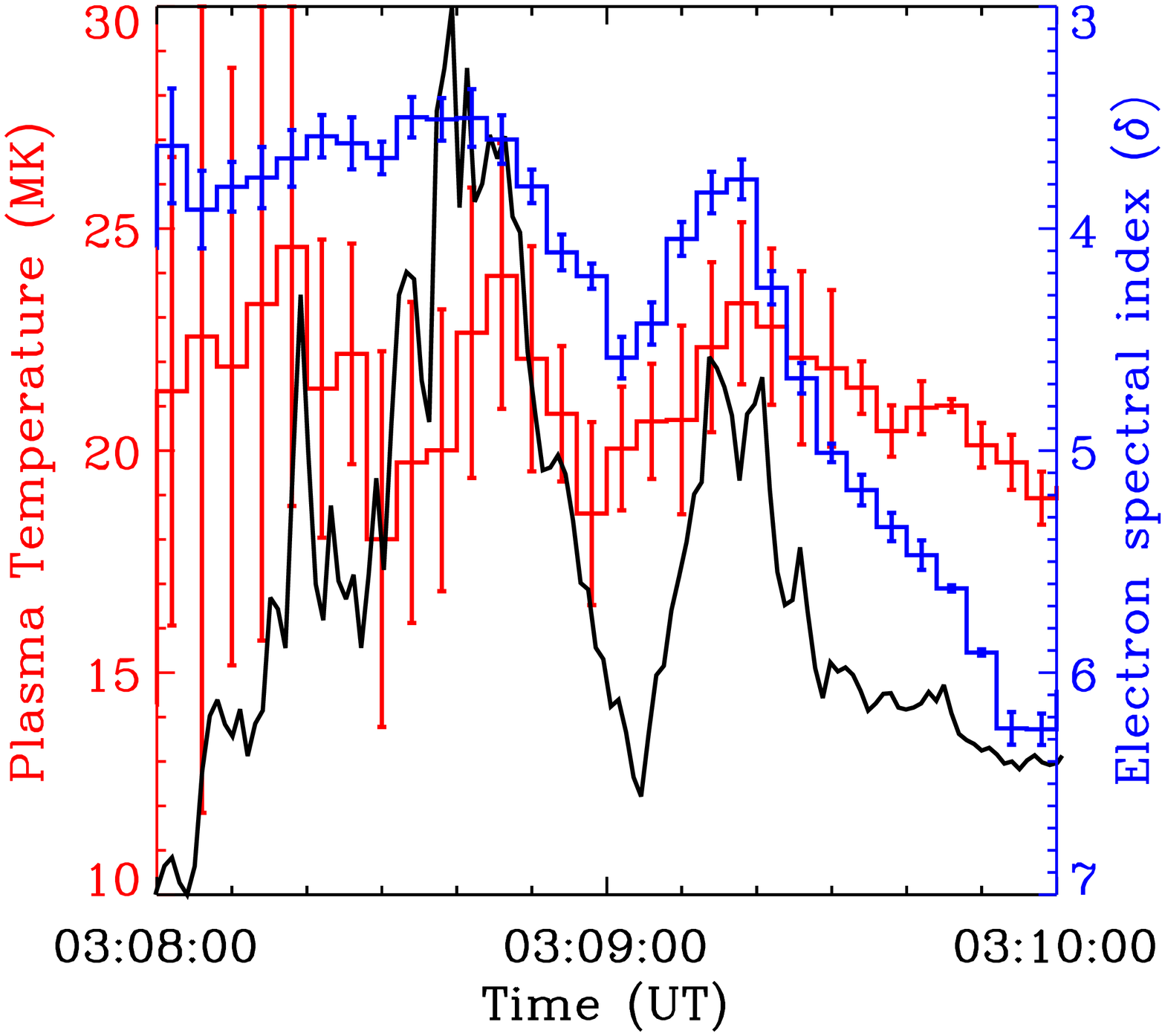}
      \includegraphics[width=0.52\textwidth,clip=]{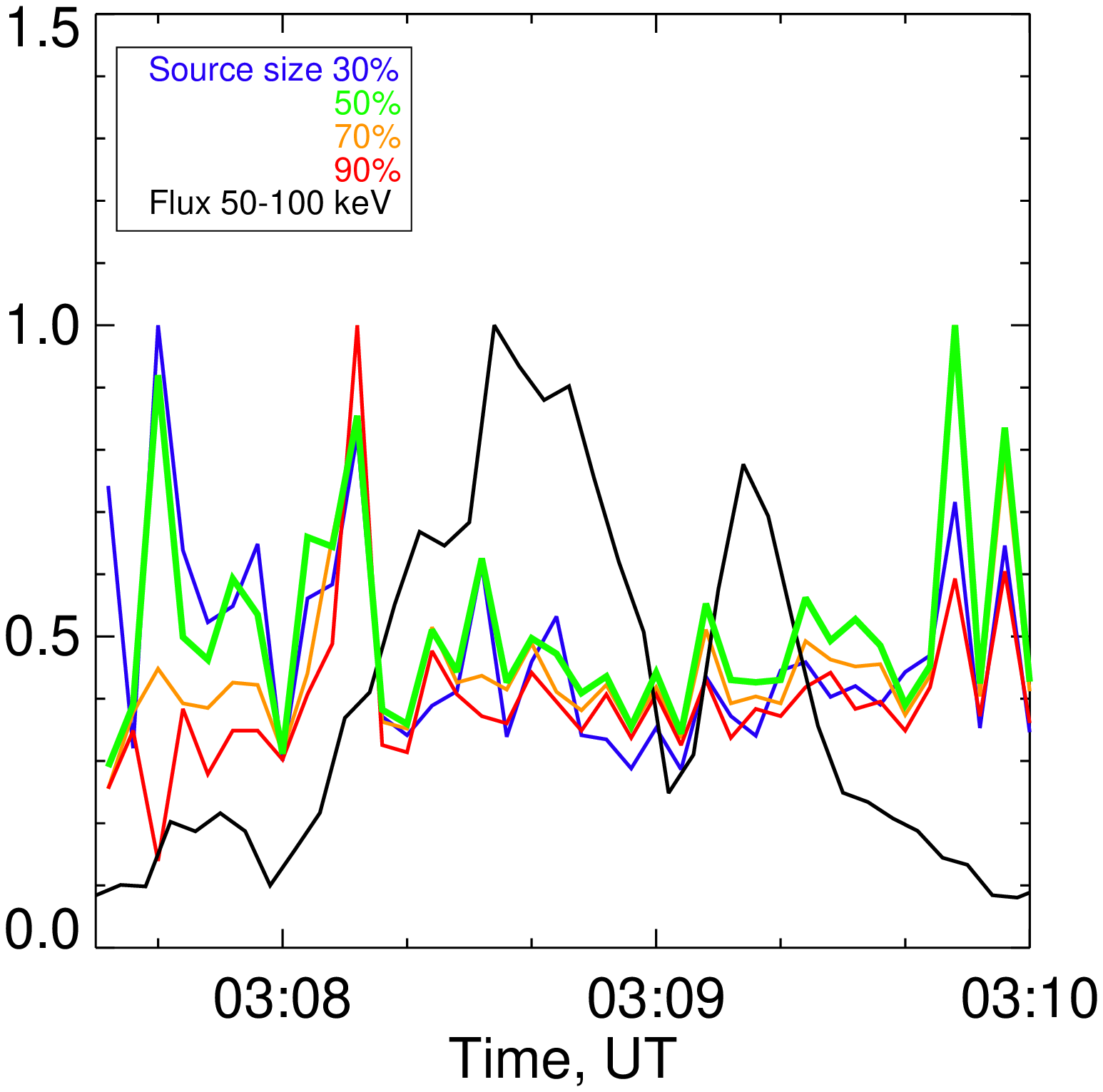}
}
              \caption{Left: time profiles of plasma temperature (red curve) and electron spectral index $\delta$ (blue curve) defined using RHESSI data with the cadence time $\Delta t =$~4~s. The error bars are the 1--sigma errors returned by the fitting routine for each fit parameter.  The black curve corresponds to the NoRH 17~GHz flux. 
Right: normalized time profiles of the source size measured within 30\% level (blue curve), 50\% level (green curve), 70\% level (orange curve), and 90\% level (red curve) of the maximal RHESSI 3--10~keV flux.}
   \label{f:T_delta}
   \end{figure}
The microwave and HXR emission of the flare demonstrate a good correlation between the data obtained by different instruments and within different energy bands (see Figure~\ref{f:5}, and also Figure~2 of Paper~1). 
 Moreover, the analysis of the spatial structure of the microwave sources reveals a good correlation of the fluxes in different areas of the flaring region (see Figure~9 and Figure~12 of 
Paper~1).
The cross-correlation coefficients are $R \approx 0.8$--0.95 for zero time delay.
The similarity of the time profiles means the observed fluxes have a solar origin, excluding any instrumental effects and any impact of the Earth's atmosphere. 

The type III bursts seen at the frequencies between 25 to 180 MHz were observed during the flare simultaneously with the microwaves and X-rays (Figure~\ref{f:RSTN}).  From \textit{Wind}/WAVES data we see that each type III burst continued to lower frequencies down to 200 kHz.
To study the relationship between the radio bursts and the microwaves/X-rays we analyse five frequency bands in the 25--180~MHz range with mean frequencies 32~MHz, 39~MHz, 53~MHz, 81~MHz, and 152~MHz. The bandwidths $\Delta f = $~3~MHz are selected for frequencies $f < 75$~MHz, and $\Delta f = $~5~MHz for frequencies $f > 75$~MHz.
 The difference in the bandwidths is caused by the difference in the frequency resolution for these two frequency bands. The bounds of each slice are marked by colored dashed lines in the left panel of Figure~\ref{f:RSTN}. 
We selected the above slices to have the minimal noise level.
For each slice, the 1D time series is calculated as a total over the bandwidth and normalized by the bandwidth. Each time profile is overplotted in the right panel of Figure~\ref{f:RSTN} after subtraction of the minimal flux.

We identify four distinct type~III bursts in the time range  03:07:00 to 03:10:30 UT (Figure~\ref{f:5}). 
The first burst with the maximum at 03:07:40~UT does not appear to be related to a peak in the microwave or HXR emission.  It occurs just before the flare impulsive phase.  However, sources of the flaring emission appear in the microwave and X-ray maps starting from 03:07:15--03:07:30~UT (Paper~1).  

The second, third, and fourth radio bursts correspond much better to peaks in the microwaves.
The second peak with the maximum at 03:08:20~UT corresponds well to that at 34~GHz.  However, it is only well pronounced at the 
lowest frequencies below 35~MHz.  At higher frequencies (35--75~MHz) only a sharp onset appears at 03:08:00--03:08:05~UT and so the burst could correspond to the smaller peak in 34~GHz 15~s earlier.

The third and fourth radio bursts at the lowest frequencies reach their maxima at 03:08:50~UT and at 03:09:55~UT respectively.
These peaks occur 10~s and 25~s after the peaks in the GHz, respectively.  The fourth peak is also quite broad in time compared to the other type~III bursts. 
As these peaks have a long rise time, an alternative method to estimate the delays between the radio and microwave/HXR bursts is to use the instants of time when the emission rises.  We estimated the delay between the rises of the bursts in microwaves and the radio bands to be 
$\approx 6\pm1.5$~s for the third radio peak and $\approx 11\pm1.5$~s for the fourth radio peak, with the uncertainty from the temporal resolution of the radio data. 

The temporal profiles of the plasma parameters were obtained as a result of fitting the HXR average spectra observed by RHESSI with a model consisting of a thermal and a thick target (based on power-law function) components. We used RHESSI software and the SPectrum EXecutive (SPEX) software \citep{2002SoPh..210..165S} for spectrum processing. The flare occurred on the limb of the Sun so we did not need to correct for albedo.
We analyzed the time profiles of the plasma temperature and electron spectral index $\delta$  within the time interval 03:07:00--03:10:00~UT with the cadence time $\Delta t =$~4~s (Figure~\ref {f:T_delta}).
Temperatures are defined with big errors until 03:08:20~UT. The errors are caused by the small soft X-ray flux, or thermal component, in the beginning of the flare. Starting from 03:07:20~UT the temperature decreases quasi-periodically from 22--23~MK to 19~MK, approximately. The quasi-periodicities have the shape of damped oscillations with the average amplitude $\Delta T / T \approx 10$\%. 

The hardest electron spectral index ($\delta \approx 3$) occurrs about 03:08:40 UT and coincides with the peak of the maximum intensity. The time profile of the electron spectral index does not show the quasi-periodic pulsations as clearly as the HXR flux, but does show the typical soft-hard-soft behavior observed during in the majority of flares.
Three peaks are pronounced near 03:08:20, 03:08:40, and 03:09:20~UT, and coincide in time with the maxima of the microwave and HXR emission  (Figure~\ref{f:5}), as well as with the maxima of the temperature (Figure~\ref {f:T_delta}, left panel). The temperature time profile also has a maximum at 03:09:45~UT.

\section{Quasi-periodic variations}\label{s:QPP}
 \begin{figure}   
   \centerline{
\includegraphics[width=0.50\textwidth,clip=]{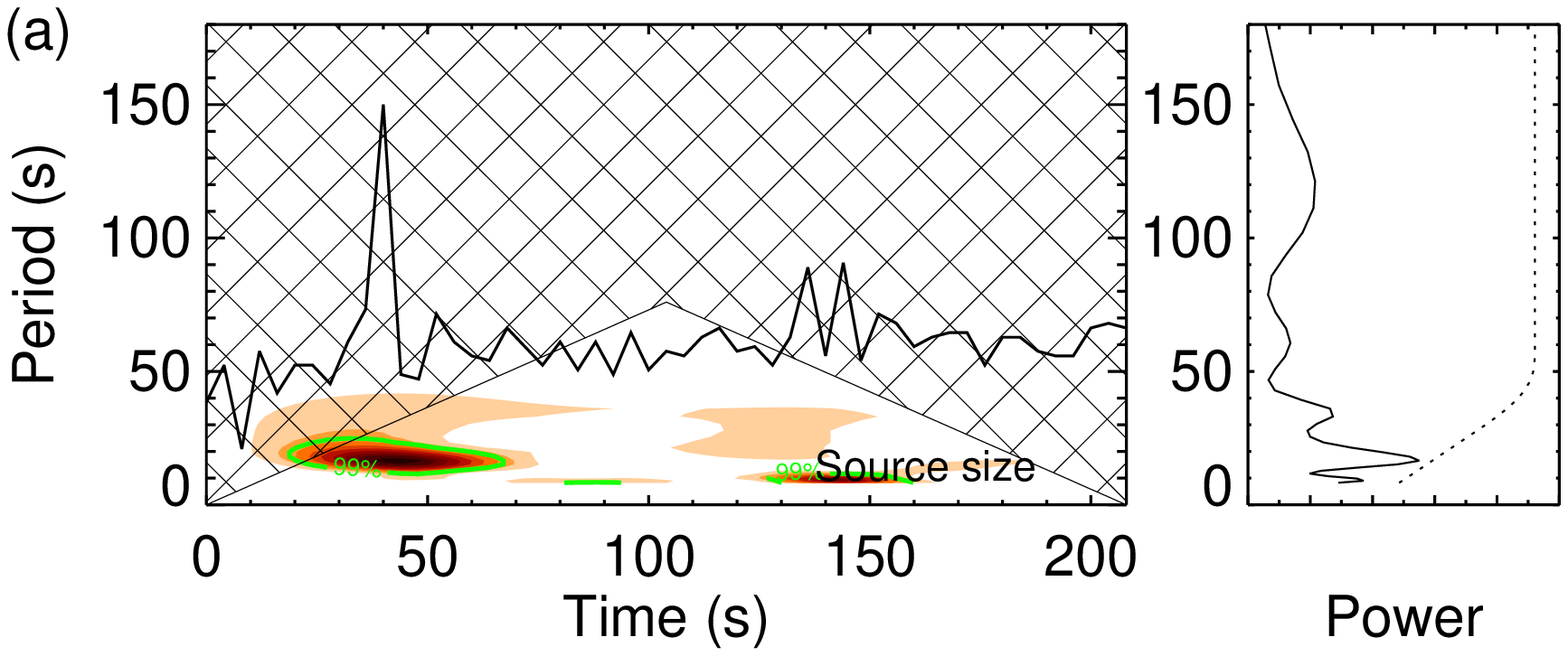}
\includegraphics[width=0.50\textwidth,clip=]{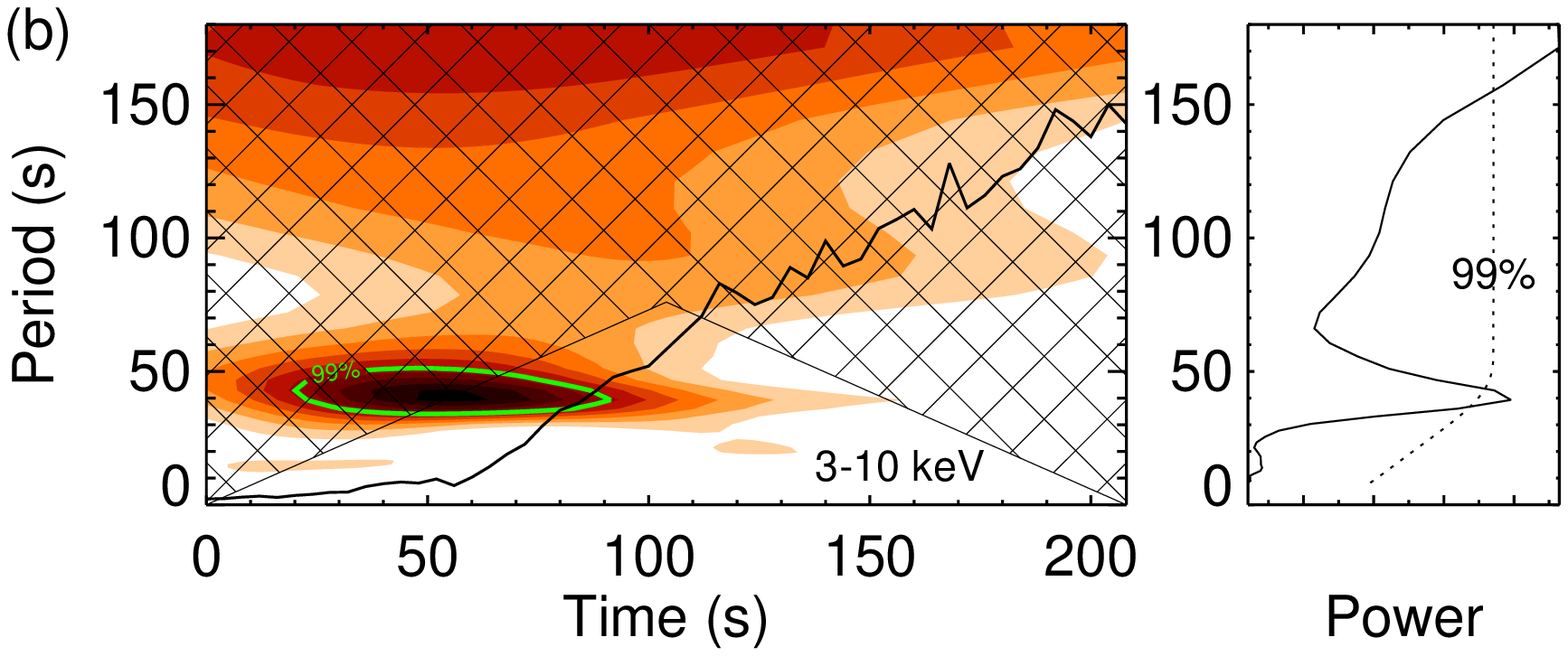}
}
   \centerline{
\includegraphics[width=0.50\textwidth,clip=]{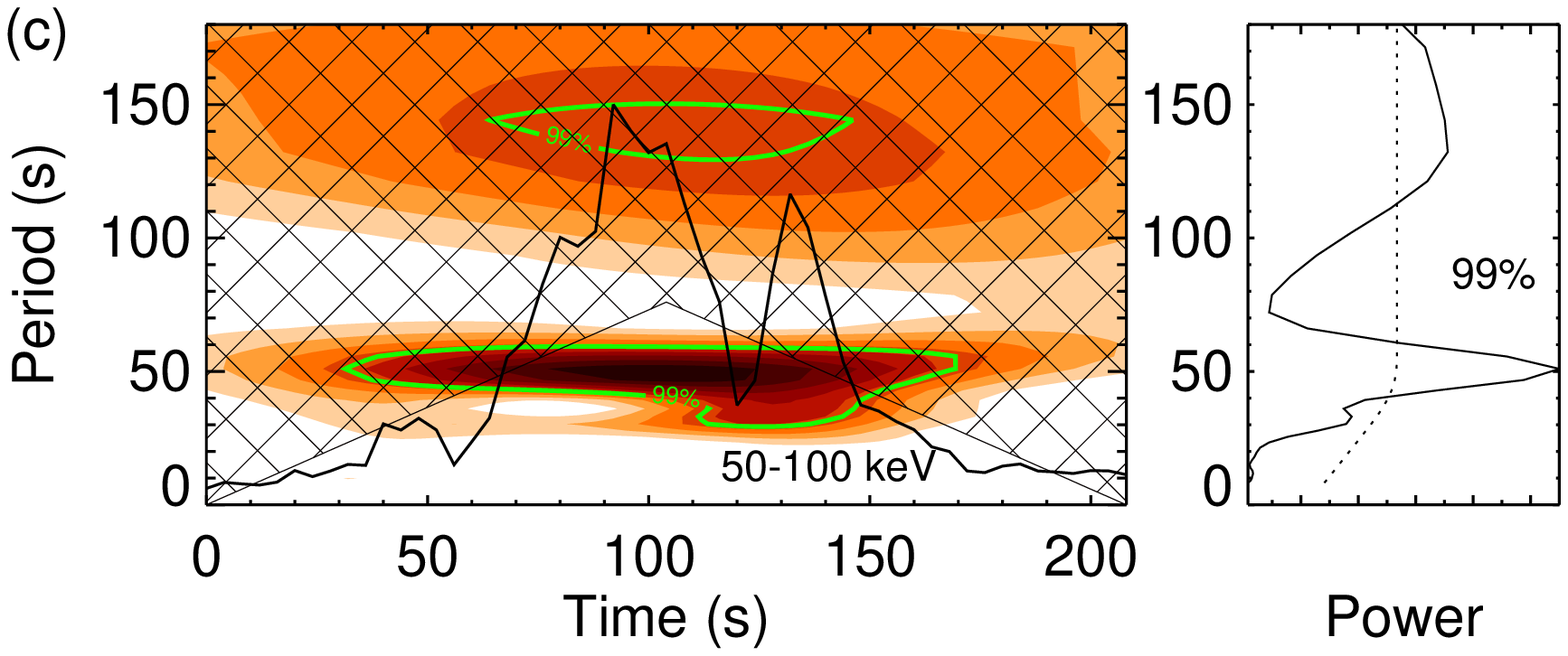}
\includegraphics[width=0.50\textwidth,clip=]{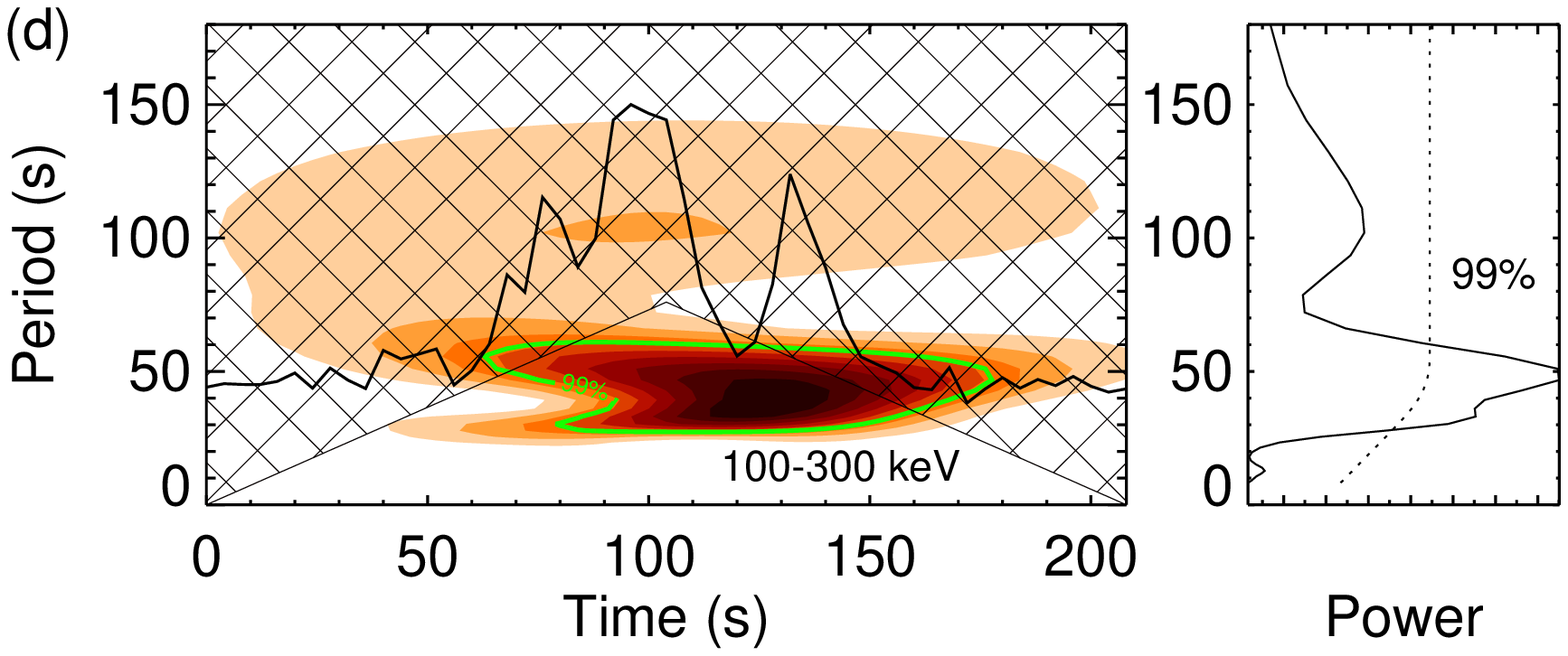}
}
   \centerline{
\includegraphics[width=0.50\textwidth,clip=]{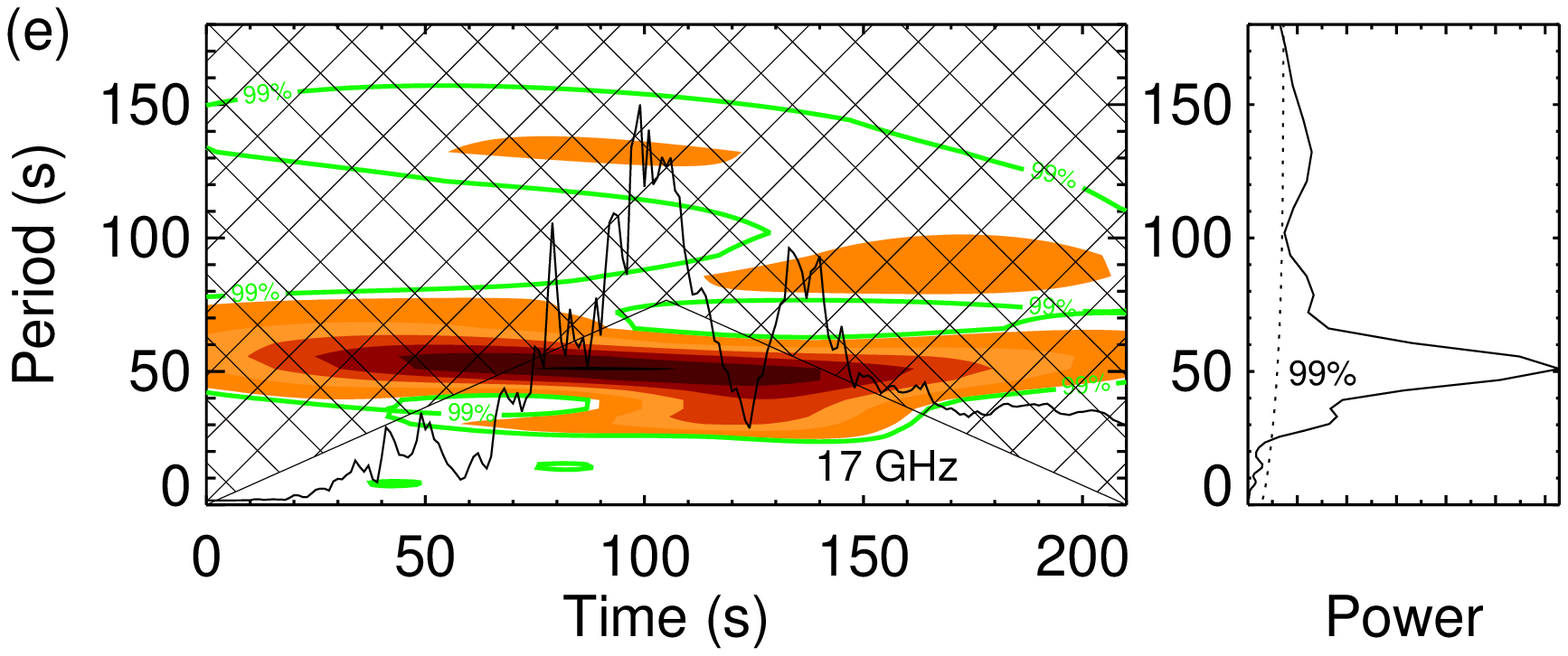}
\includegraphics[width=0.50\textwidth,clip=]{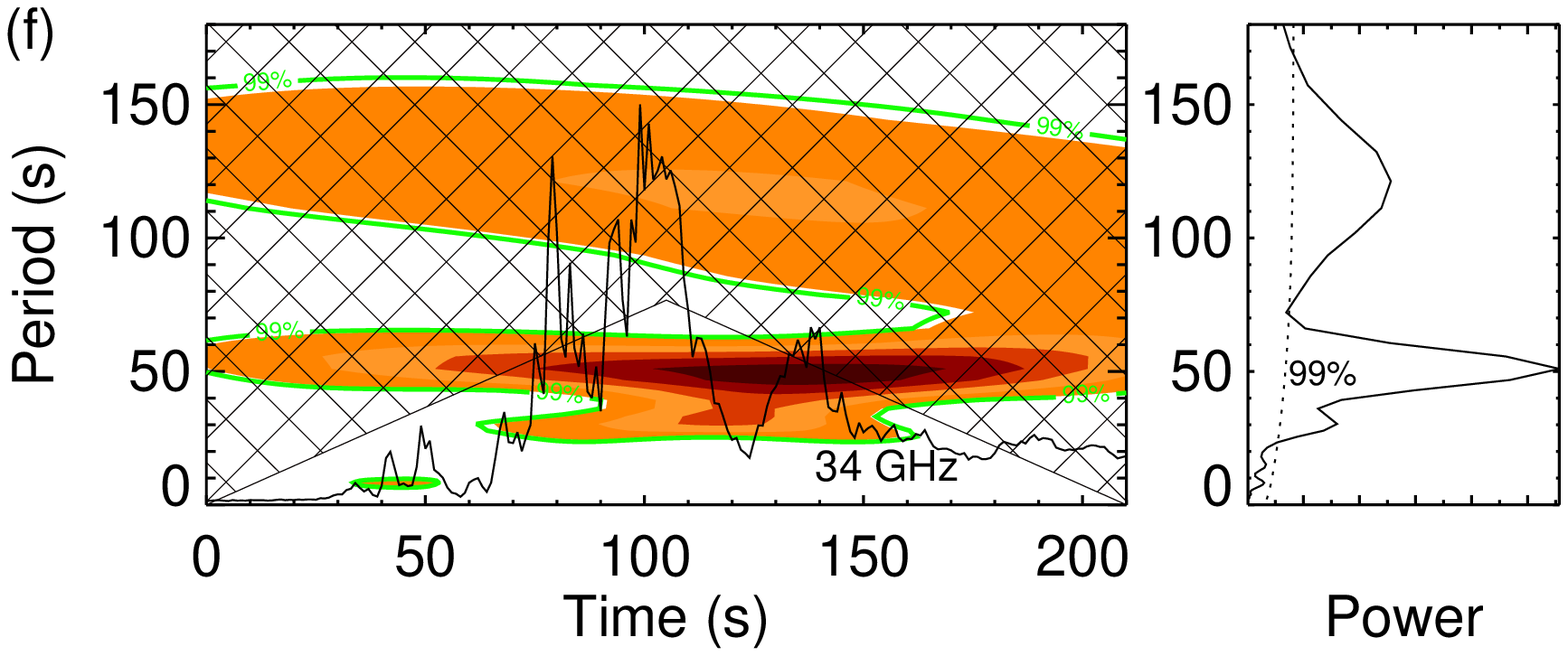}
}
   \centerline{
\includegraphics[width=0.50\textwidth,clip=]{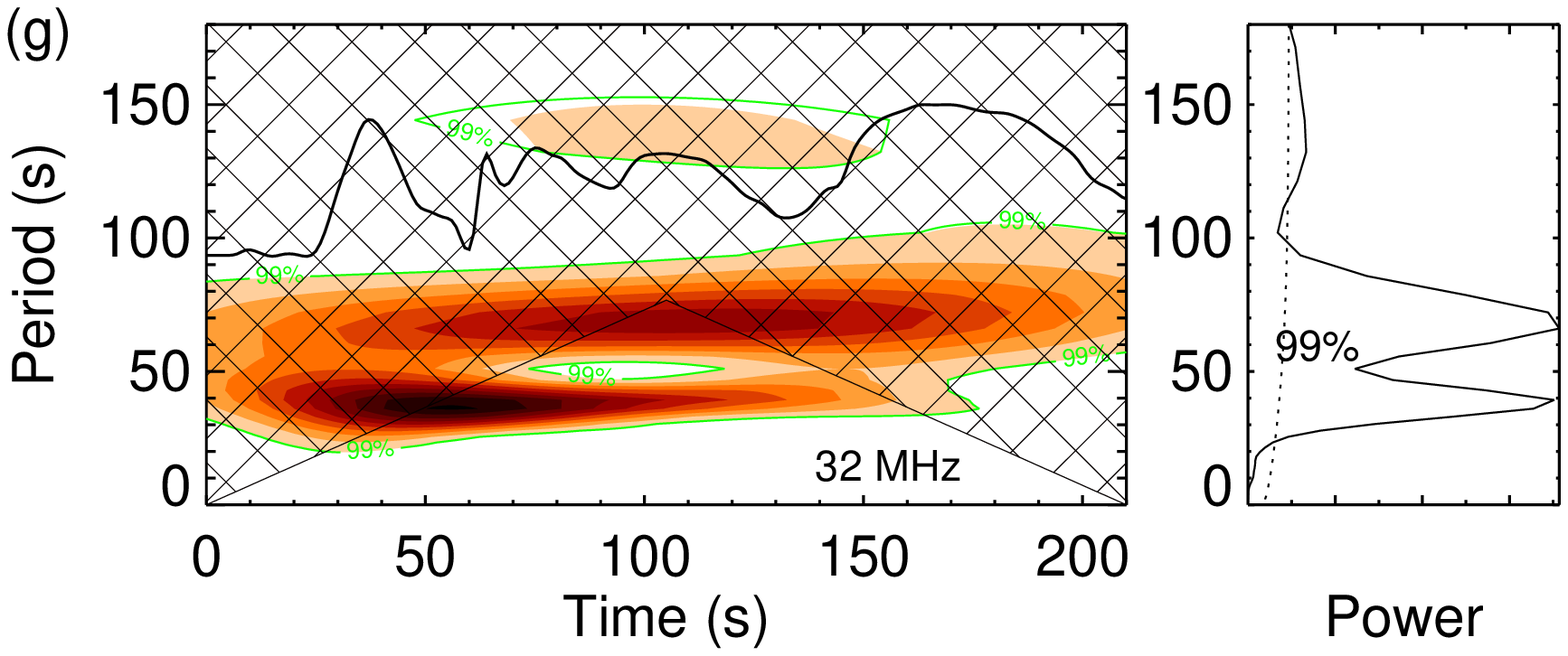}
}
   \caption{
Wavelet power spectra (colored plots) of the high-frequency component of different data. The width of the smoothing interval $\tau = 60$~s is used. In each panel, the normalized time profile is overplotted in the wavelet power spectrum. The green contour indicates the 99\% significance level. The plot to the right of the colored one is the global wavelet spectrum obtained by integration of the wavelet power spectrum over time. The dashed line here shows the 99\% significance level. The panels correspond to the following time profiles:
(a) the thermal X-ray source size calculated within 50\% level of the maximal flux at 3--10~keV, 
(b) the RHESSI flux at 3--10~keV, 
(c) the RHESSI flux at 50--100~keV, 
(d) the RHESSI flux at 100--300~keV,
(e) the NoRH flux at 17~GHz,
(f) the NoRH flux at 34~GHz,
(g) the RSTN Learmonth flux at 32~MHz.
The time axis corresponds to the interval 03:07:00--03:10:30~UT for panels from (a) to (f) 
and to 03:07:00--03:12:00~UT for panel (g).
   }
 \label{f:Wavelets}
 \end{figure}
We study the periodic properties of the flare emission using both standard wavelet routines with the Morlet base functions, and Fourier periodogram analysis. The wavelet procedure can be applied to the centred time series only. 
	We apply a method of smoothing with running average  over time $\tau$ from 30~s to 200~s
to subtract a low-frequency background from the original time series  \citep[see][for details]{2010SoPh..267..329K}. 
The residual is a high-frequency component of the signal, which constitutes a centred time series.
We note that the total amplitudes of the high-frequency component are considered without Fourier filtration \citep{2009A&A...493..259I}. Both techniques give similar results so we show in this section the wavelets only.

Fitting the RHESSI energy spectrum allows us to distinguish between thermal  ($\leq$~10 keV) and non-thermal ($\geq$~25--50~keV) components of the X-ray emission 
(Paper~1). 
We estimate the source size within different levels, from the maximal value of the flux at 3--10~keV at each time from 03:07:00 to 03:11:50~UT with $\Delta t = 4$~s. The time behavior of the source size within 30\%, 50\%, 70\%, and 90\% of 
the maximum is shown by different colored lines
in the right panel of Figure~\ref{f:T_delta}. 
All the time profiles look similar and so we choose the 50\% level time profile as the example of the source area variation.
 It is clear from the wavelet spectrum (Figure~\ref{f:Wavelets}, panel a) that there are no periodic changes in the size of the thermal source. 
The wavelet spectrum for the thermal X-ray flux reveals a very weak spot at the period of 40~s (Figure~\ref{f:Wavelets} panel b). If we look at the 99\% significance level, it is obvious that the spot covers only one or one and half periods, so this periodicity can not be considered as a QPPs.

The non-thermal emission reveals high-amplitude quasi-periodic changes with the dominant time scales of $P_1 \approx $~50~s and $P_2 \approx $~30~s. Both periods are well pronounced in the microwaves at 17~GHz and at 34~GHz (Figure~\ref{f:Wavelets} panels e and f) and HXRs at 50--100~keV and 100--300~keV bands (Figure~\ref{f:Wavelets} panels c and d), as well as at 25--50~keV (Paper~1). A similar periodicity, $P_1 \approx $~40--50~s, is also found in the flux at 32~MHz (Figure~\ref{f:Wavelets} panel g).
This confirms that the $P_1$ and $P_2$ periodicities are not simply instrumental. Besides,
three or four cycles of the oscillation are well pronounced in both the
original signals and in the high-frequency component 
for all smoothing intervals $\tau$. This allows us to exclude (or sufficiently decrease) the artificial impact of the mathematical method on the results of the spectral analysis.
The average relative amplitude of the variations in microwaves and X-rays is very high, $\Delta I / I \approx 30$\%, and it reaches 80\% after the flare maximum. 

The longer period, $P_3 \approx $~80~s variations are present in the time profile at 32~MHz  (Figure~\ref{f:Wavelets} panel g). It is attributed to two smooth peaks visible from 03:09:10~UT to 03:12:00~UT. Periods $P_4 \leq$~100~s, seen in the global wavelet spectra at 17~GHz and 34~GHz, are attributed to the global trend of the flare.

All the wavelet spectra are shown in this paper for a smoothing interval of $\tau = 60$~s.
 However, the QPPs keep their periods for all  $\tau$ from 30~s to 200~s. 
Therefore the QPPs belong to the signal 
and they are not artifacts of the smoothing. 
The results of the wavelet analysis are confirmed by the results of Fourier analysis. 
Due to the discreteness of the frequency grid in Fourier and wavelet transforms 
the precisions of the periods determined are $\Delta P_1 = \pm  4$~s 
for the period $P_1 \approx 50$~s and $\Delta P_2 = \pm  1$~s for period $P_2 \approx 30$~s.
These peaks in the Fourier periodogram are rather wide with the full width at half power 
$FWHP_1 \approx 16$~s and $FWHP_2 \approx 7$~s respectively.

\section{Discussion}\label{s:Discussion}
Analyzing the time behavior of the flare on May 6, 2005 we found common periodic variations (40--50~s) of the microwave and hard X-ray emission and in the type III radio bursts. The average relative amplitude of the variations is high at 30\% above the background flux level, reaching 80\% during the flare maximum.  However, we did not find this periodicity either in the thermal X-ray flux component or source size dynamics. 

We consider two probable explanations of QPPs in flare emission: MHD oscillations in plasma waveguides and oscillatory reconnection.  In both cases we would observe pulsations in HXRs and microwaves caused by accelerated electrons.  The MHD oscillation model explains the high amplitude of oscillations by modulation of the emission produced by accelerated electrons \citep[see][as pioneer studies]{1969ApJ...155L.117P, 1983Natur.305..292N, 1983ApJ...271..376K} whilst the oscillatory reconnection modulates the process of electron acceleration \citep{2009SSRv..149..119N}.

The wavelet analysis found a similar periodicity in the time profiles with period $P \approx 40$--$50$~s 
for all the studied spectral regions. This confirms that the observed QPPs are of solar origin and not artifacts 
due to data handling or due to instrumental effects \citep{2011A&A...530A..47I}.

\subsection{MHD oscillations}\label{s:MHD}

Microwave emission from solar flares is often attributed to gyrosynchrotron emission from accelerated electrons. The simplified equation for the gyrosynchrotron emissivity \citep{1982ApJ...259..350D} gives the relationship between the intensity $I$ and the magnetic field $B$
$$ I \propto N B^{0.9\delta - 0.22}.$$
Here, $N$ is number of accelerated elecrtons with the energies more than 10~keV. 
MHD waves do not significantly affect the variations of number of accelerated electrons. 
Thus we may assume that the number $N$ is nearly constant. 
The spectral index $\delta$ of accelerated electrons derived from the HXR observations varied from $\delta \approx 3.5$ in the beginning of the impulsive phase to $\delta \approx 5.5$ at the end of the impulsive phase (Figure~\ref{f:T_delta}).  If we assume the angle between the magnetic field and the line of sight $\theta = const$ then the dependence of the intensity on the magnetic field is roughly estimated as $I \propto B^3$ for the harder spectrum with $\delta \approx 3.5$. Taking the logarithmic derivative from this relationship we calculate $\Delta B/B = \Delta I/3I$. The average amplitude of the QPPs found in Section~\ref{s:QPP} is $\Delta I / I \approx 30$\% so we obtain $\Delta B/B \approx 10$\%. 
In a similar way we estimate the variations of the magnetic field for the softer spectrum at the end of the impulsive phase. The spectral index $\delta \approx 5.5$ leads to  $\Delta B/B = \Delta I/5I \approx 6$\%.

The amplitude of the QPPs reaches $\Delta I / I \approx 80$\% at the flare maximum where $\delta \approx 4.5$. In this case the variations of the magnetic field should be $\Delta B/B \approx 20$\% in order to provide such a high modulation depth.  It is hard to explain this property in terms of the modulation of the emission by MHD oscillations of the emitting volume. 

Another argument against the production of the detected QPP by direct modulation of the emitting plasma or magnetic field by an MHD wave is the absence of pronounced oscillations in the thermal X-ray fluxes or the thermal X-ray source size dynamic  (Figure~\ref{f:Wavelets} panels a and b). 

We found the time profiles of the microwaves, HXRs and type III radio bursts varying quasi-periodically, with a common period of 40--50 s.   Speculatively, such stable periodicity within so wide a height range is difficult to explain by MHD oscillations.  Moreover, it is not obvious how an MHD oscillation would affect the amplitude of type III emission to cause the observed periodicity.  Type III bursts are believed to be caused by discrete electron beams traveling around $10^{10}$~cm~s$^{-1}$, much faster than typical speeds of MHD waves.

\subsection{Quasi-periodic reconnections}\label{s:Reconnections}

The good correlation between the time profiles in microwaves and HXRs  (Figure~\ref{f:5})  suggests that both emissions are generated by the same, time-varying population of accelerated electrons.  This fact allows us to consider periodic reconnection as a possible scenario of observed 
variations of the flare emission intensity.
\citet{2015ApJ...807...72L} found QPPs with $P \approx $~4~min in HXRs, in the chromospheric and coronal line emissions, and in the radio emission. The authors showed that each radio peak corresponds to a type III burst. The results of that study indicate that the QPPs observed in this flare were generated by the non-thermal electrons accelerated by a set of quasi-periodic magnetic reconnections.

As it was shown in Section~\ref{s:Timeprofs} the type III bursts in the radio MHz band look like they are related to the microwaves and the HXRs (Figure~\ref{f:5}). The relative phase relationships between these emissions could indicate that electrons are accelerated in the corona giving rise both to the development of the microwave and X-ray sources in lower layers, as well as to the type~III bursts in the higher layers of the solar corona. Moreover, we found the $P_1 \approx $~40--50~s periodicity which is similar to the periodicity detected in HXR and microwave emission (Figure~\ref{f:Wavelets} panels c to g). 

The time evolution of the electron spectral index $\delta$ calculated using RHESSI data is characterized by the presence of three peaks (Figure~\ref{f:T_delta}) in the spectral hardness, near the times 03:08:20, 03:08:40, and 03:09:20~UT.  These peaks coincide in time with the peaks of microwave emission at 17~GHz (Figure~\ref{f:5}).  When the microwave flux increases, the spectral index decreases and the spectrum becomes harder, implying progressive acceleration of electrons. Similarly, the decrease in the microwave flux and simultaneous increase in $\delta$ reveal the relaxation of the acceleration process. 

The second, third and fourth type III bursts appear to be related to the bursts observed in microwaves and HXRs.  There is an expected time delay between a peak in microwaves and a corresponding peak at 32 MHz, related to the travel time of the electrons from the acceleration region to the upper corona.  Assuming a density model \citep{1958ApJ...128..677P} of the quiet Sun, we attribute 32~MHz to around $5.4 \times 10^{10}$~cm.  A modest type~III speed of 0.2~$c$ (where $c$ is the light speed) gives just under 10~s travel time that is similar to the delay found between the rise in microwaves and the rise in type III emission (Section \ref{s:Timeprofs}).  This provides further evidence for a common acceleration region responsible for the energetic electrons that created the type III emission, microwaves and HXRs.

\section{Conclusion and final remarks}\label{s:Conclusion}
Analyzing the microwaves, hard X-rays and type III bursts emitted during a solar flare we found the following results: a stable 40--50~s periodicity from the microwaves, hard X-rays and type III radio bursts, a similar periodic behavior of the electron spectral index, and an absence of any periodicity in the thermal X-ray lightcurves or the thermal X-ray source size.  We conclude that the observed QPPs were most probably caused by quasi-periodic acceleration of electrons caused by oscillations in the current sheet rather than high-amplitude MHD pulsations. 

This work adds to the other confirmations from observations \citep{2015ApJ...804....4K} and from simulations  \citep{2009A&A...494..329M, 2012ApJ...749...30M} that magnetic reconnection can be modulated by MHD waves with minute periods.  
The oscillations in the current sheet could be caused by slow magneto-acoustic waves coming from the lower layers of the solar atmosphere.  The indirect evidence in favor of MHD driven reconnections could be the slow magneto-acoustic waves found during the decay phase of this flare with an approximately similar period (Paper~1).
This suggestion is in agreement with the results of simulations made by \citet{2012A&A...548A..98M} 
who considered non-linear fast magnetoacoustic waves that deform a magnetic X-point as a driver for oscillatory reconnection with periods from 56.3 s to 78.9 s.
However, diagnostics of the wave type is impeded because of the absence of EUV data with high temporal resolution.


 \begin{acks}
This work was (partly) carried out on the Solar Data Analysis System 
operated by the Astronomy Data Center in cooperation with 
the International Consortium for the Continued Operation of Nobeyama Radioheliograph (ICCON).
This research was partly supported by the grants of the Russian Foundation 
for Basic Research  No.No. 14-02-00924, 15-02-08028, 15-02-03717, 
by the Program of Russian Academy of Sciences No.22, 
by the RAS Presidium program No. 0344-2015-0017, 
by the Marie Curie FP7-PEOPLE-2011-IRSES-295272, 
and by STFC consolidated grant ST/L000741/1 (HASR).
EK is a beneficiary of a mobility grant from the Belgian Federal Science Policy Office.
Authors thank an anonumous referee for helpful comments on the manusctript.
 \end{acks}

%
%
 \bibliographystyle{spr-mp-sola}
\bibliography{refs_KKRM} %
%
%
%

\end{article} 
\end{document}